\documentclass[pdflatex,sn-nature]{sn-jnl}

\usepackage{graphicx}%
\usepackage{multirow}%
\usepackage{amsmath,amssymb,amsfonts}%
\usepackage{amsthm}%
\usepackage{mathrsfs}%
\usepackage[title]{appendix}%
\usepackage{xcolor}%
\usepackage{textcomp}%
\usepackage{manyfoot}%
\usepackage{booktabs}%
\usepackage{algorithm}%
\usepackage{algorithmicx}%
\usepackage{algpseudocode}%
\usepackage{listings}%
\usepackage{setspace}
\theoremstyle{thmstyleone}%
\theoremstyle{thmstyletwo}%

\theoremstyle{thmstylethree}%

\begin{document}

\title[Article Title]{Nonradiative Multiphonon Model of Deep-Level Transient Spectroscopy: Beyond Henry-Lang Model}

\author*[1]{\fnm{Menglin} \sur{Huang}}\email{menglinhuang@fudan.edu.cn}
\equalcont{These authors contributed equally to this work.}

\author[1]{\fnm{Shanshan} \sur{Wang}}
\equalcont{These authors contributed equally to this work.}

\author[1]{\fnm{Junjie} \sur{Zhou}}

\author[1]{\fnm{Xinjing} \sur{Guo}}

\author*[1]{\fnm{Shiyou} \sur{Chen}}\email{chensy@fudan.edu.cn}

\affil[1]{\orgdiv{College of Integrated Circuits and Micro-Nano Electronics, and Key Laboratory of Computational Physical Sciences (MOE)}, \orgname{Fudan University}, \orgaddress{\city{Shanghai}, \postcode{200433}, \country{China}}}


\abstract{Deep-level transient spectroscopy (DLTS) is a key experimental method for defect characterization, yet its analysis remains controversial, and the two widely used models developed by Henry and Lang are conflicting. We show that the Henry-Lang model is valid only under the Condon approximation, as well as high-temperature and strong electron-phonon coupling approximations, which cause incorrect temperature dependence of carrier emission and capture. Here we develop a rigorous nonradiative multiphonon (NMP) model, and demonstrate that the temperature dependence is governed predominantly by effective phonons with large phonon wavefunction overlap and high thermal occupation. The effective phonons are strongly correlated with lattice relaxation.
The neglect of this correlation in existing DLTS models introduces substantial errors when they are used to fit DLTS-measured emission rates. Our comparison for 21 different defects in 12 semiconductors, including Si, SiC and Ga$_2$O$_3$, shows that the Henry-Lang model gives a completely different temperature dependence of carrier capture cross section from that obtained using the rigorous NMP model, with errors reaching
up to six orders of magnitude at room temperature. Our study highlights the necessity of revisiting previous DLTS analysis studies using the rigorous NMP model.

}




\maketitle
\section{Introduction}\label{sec1}

\par Electrically active defects in semiconductors control carrier recombination, transport, trapping, leakage, and long-term reliability of devices \cite{science}. Establishing their microscopic energetics and understanding their interaction with carriers are therefore important to semiconductor physics and device engineering. Deep-level transient spectroscopy (DLTS), since its invention in 1974 \cite{Lang1974JAP}, has acted as the most important experimental technique for probing energy levels, concentrations, and carrier capture properties of defects in semiconductors \cite{kim2021NP,zhao2023NC,Wang2025JPAP,liu2025NC}. For example, DLTS has been used to identify deep levels across the band gap of $\beta$-Ga$_2$O$_3$, and helps clarify the microscopic mechanisms of its leakage and breakdown behaviors in power electronic devices \cite{Zhang2016APL,Fregolent2024JPD}.

\par
\begin{figure}[htbp]
	\centering
	\includegraphics[width=0.9\textwidth]{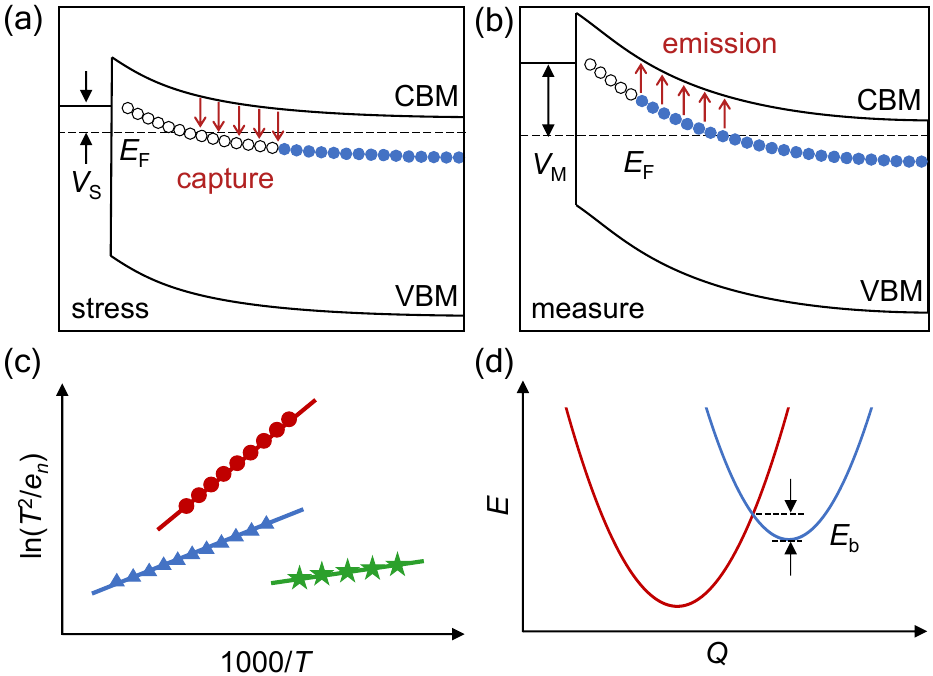}
	\caption{(a, b) Schematic band diagram during stressed and measuring periods in a DLTS measurement. (c) Linear fitting of $\ln(T^2/e_n)$ vs $1/T$. (d) Configuration coordinate diagram of an NMP transition with a barrier from initial to final state.}
	\label{fig1}
\end{figure}

 In DLTS measurement, a stress voltage is applied to a p-n junction, Schottky junction, or MOS structure to induce carrier capture by defects. After the stress is removed, the subsequent carrier emission process is monitored, and the emission rate $e_n$ is recorded [Fig. \ref{fig1}(a, b)]. Repeating this procedure over a range of temperatures $T$ yields the $T$ dependence of $e_n$ \cite{Khan2015Book,schroder2015semiconductor}. Through fitting the measured $\ln(T^2/e_n)$ vs. $1000/T$ data [Fig. \ref{fig1}(c)] to a theoretical model, one can obtain the defect energy level $E_T$ and carrier capture cross section $\sigma_n$ \cite{Fregolent2024JPD}. However, there are currently two conflicting models.

\par The first model was introduced by Lang in the original paper that invented DLTS \cite{Lang1974JAP}, where the carrier capture cross section $\sigma_n$ was assumed to be a constant. This leads to the relation,
\begin{equation}\label{Lang_model}
    \ln(T^2/e_n)=\frac{E_\text{C}-E_\text{T}}{k_\text{B} T}-\ln(\sigma_n \gamma),
\end{equation}
where $E_\text{T}$ is the defect level referenced to the band edge $E_\text{C}$, and $\gamma$ is a temperature-independent prefactor \cite{Deng2024IFM}. A linear fit of $\ln(T^2/e_n)$ vs. $1000/T$ [Fig. \ref{fig1}(c)] gives $E_\text{T}$ and $\sigma_n$. Owing to its simplicity, the constant-$\sigma_n$ model is still the most widely used in analyzing DLTS data nowadays.

The second model was developed by Henry and Lang in 1977 \cite{henry1977nonradiative}. They described carrier capture as a nonradiative multiphonon (NMP) process and introduced a capture barrier $E_\text{b}$ in a configuration coordinate diagram [Fig. \ref{fig1}(d)] \cite{henry1977nonradiative,Passler1978,ridley1978multiphonon,Alkauskas2016JAP1, Vasilev2024JAP}. Based on the Condon approximation, they
derived a temperature-dependent $\sigma_n$, $\sigma_n^{\text{HL}}(T)= T^{-1}\sigma_{n0} \exp \left(-E_\text{b}/k_\text{B}T \right)$, where $\sigma_{n0}$ is the so-called apparent carrier capture cross section \cite{Deng2024IFM}. Then,
\begin{equation}\label{Arrhenius}
    \ln(T^2/e_n)=\frac{E_\text{C}-E_\text{T}+E_\text{b}}{k_\text{B} T}-\ln(T^{-1}\sigma_{n0} \gamma).
\end{equation}
Here, $\ln(T^2/e_n)$ no longer follows a linear relation with $T^{-1}$. To restore linearity, the Henry-Lang model approximated the factor $T^{-1}$ in the last term as a constant \cite{henry1977nonradiative, Stampfer2024, Lan2025JAP}. Then a linear fit can give $E_\text{C}-E_\text{T}+E_\text{b}$ and $\sigma_{n0}$. The Henry-Lang model is also used in a large number of DLTS studies \cite{Vasilev2024JAP}.

The two models conflict obviously. The first model assumes $\sigma_n$ is a constant, while the second model shows an obvious temperature dependence of $\sigma_n$. When they are used to analyze DLTS-measured $e_n$ vs. $T$ data, the extracted $E_T$ and $\sigma_n$ can differ significantly, causing controversies about the defect properties. It is an open question whether the two models are correct and can be used to analyze DLTS experiments unambiguously, imposing a serious limit to the reliability of previous DTLS studies.


\section{Results}\label{sec2}
\subsection{Henry-Lang model vs. rigorous NMP model}\label{subsec2-1}

\par To answer this question, we first discuss the approximations underlying the two models. The first model simply assumes that $\sigma_n$ is constant, but there is no proof to justify this assumption. The Henry-Lang model considers the carrier emission or capture as an NMP transition process, implying that $\sigma_n$ should be temperature dependent. In the Supporting Information, we present the derivation of the temperature dependence of $\sigma_n^{\text{HL}}(T)$ in the Henry-Lang model, and clarify that it relies on the Condon approximation, high-temperature and strong electron-phonon coupling approximations. Since the late 1970s, the development of NMP theory has shown clearly that the Condon approximation is not accurate for describing NMP transitions, because the Herzberg-Teller type electron-phonon coupling can be significant \cite{Shi2012PRL,Wickramaratne2018APL,ourNMP}. Therefore, the Henry-Lang model, based on the Condon approximation, should not be expected to describe carrier emission and capture processes accurately. Moreover, the high-temperature and strong-coupling approximations are not necessarily satisfied in DLTS measurements, so the Henry-Lang model may not provide a general description for diverse defect types, especially those measured at low and moderate temperatures or involving weak coupling. In short, the correctness of the Henry-Lang model is not justified either.

These questionable approximations in the Henry-Lang model motivate us to develop a more rigorous NMP-based model. According to the NMP transition-rate formula derived from Fermi's golden rule \cite{Huang1981english,Shi2012PRL,Wickramaratne2018APL,ourNMP,Kavanagh2025EES}, $\sigma_n$ can be written as,
\begin{equation}\label{multiphonon}
\begin{aligned}
    \sigma_n^{\text{NMP}}(T)= & \frac{2\pi \Omega}{\hbar v_\text{th}} \sum_{m,n} \rho_m \left| \left<\psi_i \chi_m |H| \psi_f \chi_n \right> \right|^2 \delta(E_{im}-E_{fn})\\
    = & \frac{2\pi \Omega}{\hbar v_\text{th}} |W_{if}|^2 G(T)\\
\end{aligned}
\end{equation}
where $G(T)=\sum_{m,n} \rho_m(T) \left| \left<\chi_m |Q| \chi_n \right> \right|^2 \delta(E_{im}-E_{fn})$ is the lineshape function \cite{ourNMP}. $W_{if}=\left< \psi_i|{\partial H}/{\partial Q}|\psi_f \right>$ is the electronic coupling. $\Omega$ is the volume, $v_\text{th}=\sqrt{3k_\text{B}T/m^\star}$ is the carrier thermal velocity 
and $m^\star$ is the effective mass, $\rho_m$ is the phonon occupation of state $m$, $\psi_i$ and $\psi_f$ are the electronic wavefunctions of the initial and final states, and $\chi_m$ and $\chi_n$ are the $m$th and $n$th phonon wavefunctions with a displacement $\Delta Q$ along the configuration coordinate. The $\delta$ enforces energy conservation.

The $T$ dependence of $\sigma_n$ is reflected in $v_\text{th}$ and $G(T)$. Using the relation $\frac{d \rho_m}{dT} = \rho_m \frac{d\ln \rho_m}{dT}$, we can obtain,
\begin{equation}\label{MP_first}
\begin{aligned}
    \frac{d\ln G(T)}{dT} = & \frac{\sum_m \left[ \frac{d\ln \rho_m}{dT} \right] \rho_m |\left< \chi_m|Q|\chi_n \right>|^2}{\sum_m \rho_m |\left< \chi_m|Q|\chi_n \right>|^2} \\
    = & \frac{ \bar{E}_\text{eff}(T) -E_{\text{all}}(T)}{k_\text{B}T^2}.
\end{aligned}
\end{equation}
Here we specify the final phonon state $n\approx m+\Delta E/\hbar\omega$ with a finite broadening used in numerical evaluations, which replaces the $\delta$ function. $E_{\text{all}}(T)=\frac{\hbar \omega}{2} \coth(\frac{\hbar\omega}{2k_BT})$ is the quantum-statistical average energy of a harmonic oscillator, and $\bar{E}_\text{eff}(T)$ is an averaged energy of initial-state phonons at temperature $T$,
\begin{equation}\label{average_phonon}
    \bar{E}_\text{eff}(T) = \frac{\sum_m (m+1/2) \hbar \omega \rho_m L_m}{\sum_m \rho_m L_m},
\end{equation}
where the thermal occupation $\rho_m$ and the wavefunction overlap factor $L_m=|\left<\chi_m|Q|\chi_n\right>|^2$ jointly determine the weight of the average. We call $\bar{E}_\text{eff}(T)$ as the effective initial-state phonon energy. Based on Eqs. (\ref{multiphonon}-\ref{average_phonon}),
\begin{equation}\label{MP_derivative}
    \frac{d\ln \sigma_n^{\text{NMP}}}{dT}=\frac{\bar{E}_\text{eff}(T) -E_{\text{all}}(T)}{k_\text{B}T^2}-\frac{1}{2T}.
\end{equation}
We can thus obtain the NMP-based $\sigma_n$ formula by integrating Eq. (\ref{MP_derivative}),
\begin{equation}\label{MP_integration}
    \sigma_n^{\text{NMP}}(T)=\sqrt{\frac{T_0}{T}} \sigma_n^{\text{NMP}}(T_0)\exp \left[ \int_{T_0}^{T}\frac{\bar{E}_\text{eff}(T') -E_{\text{all}}(T')}{k_\text{B}T'^2}dT' \right],
\end{equation}
where $\sigma_n^{\text{NMP}}(T_0)$ is the capture cross section evaluated at the reference temperature $T_0$ (see the SI for more details). Apparently, Eq. (\ref{MP_integration}) of the NMP model differs from $\sigma_n^{\text{HL}}(T)$ of the Henry-Lang model. This can be more clearly seen in the temperature derivatives of $\sigma_n$. The derivative in the NMP model is mainly determined by $\bar{E}_\text{eff}(T) -E_{\text{all}}(T)$ as shown in Eq. (\ref{MP_derivative}). The derivative in the Henry-Lang model is,
\begin{equation}\label{HL_derivative}
    \frac{d\ln \sigma_n^{\text{HL}}}{dT}=\frac{E_\text{b} -k_\text{B}T/2}{k_\text{B}T^2}-\frac{1}{2T}.
\end{equation}
which shows that the derivative is determined by a $T$-independent term $E_\text{b}$ and a linear term $k_\text{B}T/2$. Therefore, the main difference in the derivatives lies in the $\bar{E}_\text{eff}(T) -E_{\text{all}}(T)$ term of the NMP model and the $E_\text{b} -k_\text{B}T/2$ term of the Henry-Lang model.

\begin{figure}[htbp]
	\centering
	\includegraphics[width=0.9\textwidth]{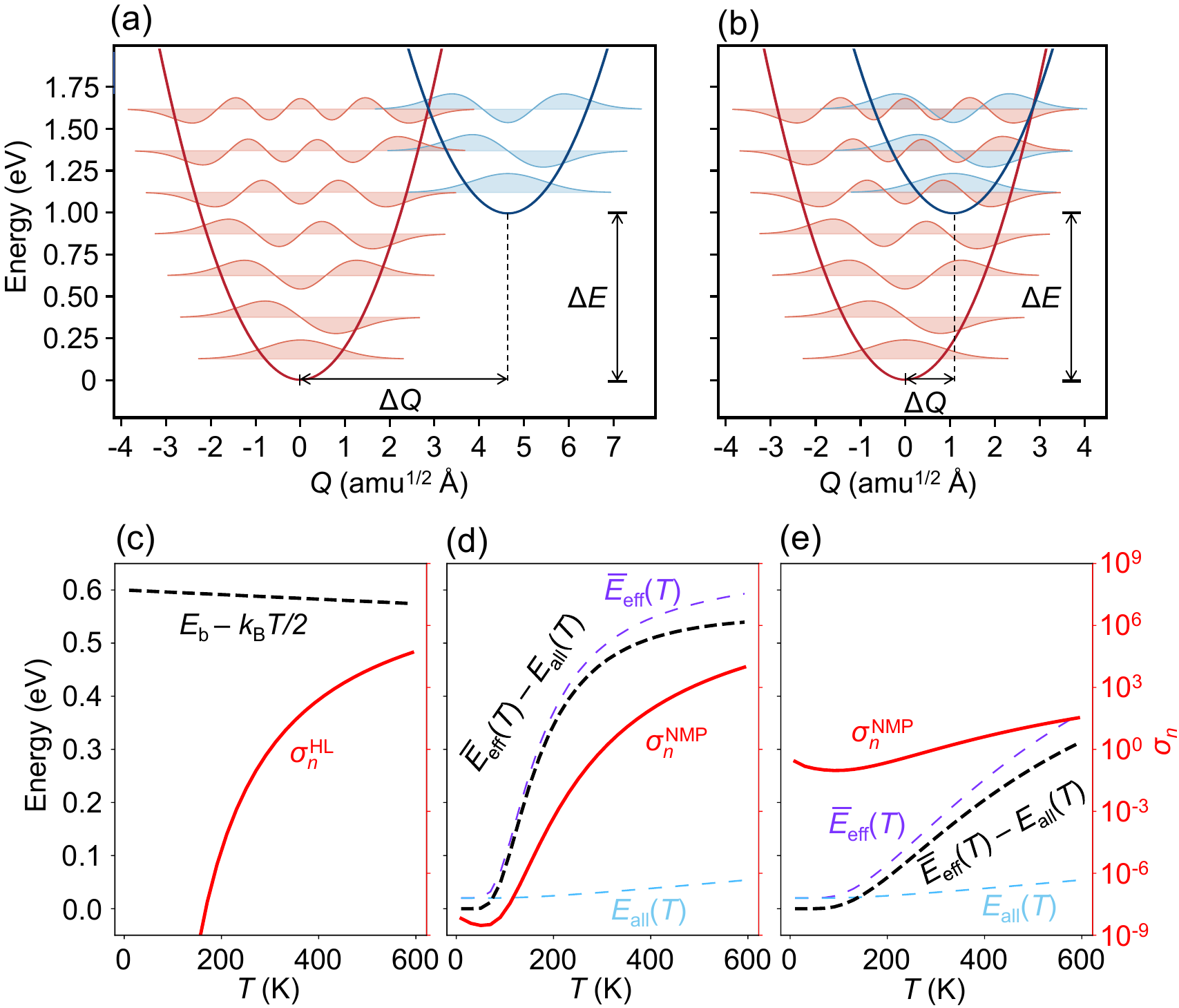}
	\caption{Configuration coordinate diagram of an NMP transition from an initial state (blue) to a final state (red) for, (a) a defect with a large $\Delta Q=4.66$ amu$^{1/2}$ $\text{\AA}$, and (b) a defect with a small $\Delta Q=1.12$ amu$^{1/2}$ $\text{\AA}$. $\omega=40$ meV and $\Delta E=1.0$ eV are same for them, which leads to the same classical barrier $E_\text{b}=0.6$ eV. The phonon wavefunction is shown by the shaded area. (c, d, e) Calculated $E_\text{b}-k_\text{B}T/2$, $\bar{E}_\text{eff}(T)-E_\text{all}(T)$, $\bar{E}_\text{eff}(T)$, and $E_\text{all}(T)$ (left axis), and $\sigma_n$ normalized at 300 K (right axis) as functions of $T$ for (c) both defects using the Henry-Lang model and (d) large-$\Delta Q$ defect, (e) small-$\Delta Q$ defect using the NMP model.}
	\label{fig2}
\end{figure}

In the following, we use two examples to show the difference between our NMP model and the Henry-Lang model. Fig. \ref{fig2}(a) shows a defect system with a phonon frequency $\omega=40$ meV, electronic transition energy $\Delta E=1.0$ eV, lattice relaxation $\Delta Q=4.66$ amu$^{1/2}$ $\text{\AA}$, and transition barrier $E_\text{b}=0.6$ eV. Fig. \ref{fig2}(b) shows another defect system with the same $\omega$, $\Delta E$, $E_\text{b}$, but with a much smaller $\Delta Q$, corresponding to weaker lattice relaxation during carrier capture and emission. According to Eq. (\ref{HL_derivative}) of the Henry-Lang model, the temperature dependence of $\sigma_n^\text{HL}$ for both defects is determined by $E_\text{b}-k_\text{B}T/2$. Since two defects have same $E_\text{b}$, the calculated $E_\text{b}-k_\text{B}T/2$ are identical. When the Henry-Lang model is used for analyzing the temperature dependence, the resulting $\sigma_n^{\text{HL}}$ are exactly same for the two defects in Fig. \ref{fig2}(c). However, in our NMP model, the temperature dependence is determined by $\bar{E}_\text{eff}(T) -E_{\text{all}}(T)$, which differs substantially between the two defects in Fig. \ref{fig2}(d, e). Consequently, the resulting temperature dependence of $\sigma_n^{\text{NMP}}$ differs significantly for the two defects in Fig.~\ref{fig2}(d, e).

We now analyze why our NMP model predicts markedly different behavior for the two defects. For the first defect, $\Delta Q$ is large, so the two potential energy surfaces (PESs) are well separated and intersect between their minima, corresponding to the Marcus normal region \cite{Marcus1956JCP}. The phonon wavefunctions of the initial state (blue shaded area in Fig.~\ref{fig2}(a)) have almost no overlap with those of the final state (red shaded area in Fig.~\ref{fig2}(a)) for low-energy phonons (small $m$). Significant overlap appears only for high-energy phonons (large $m$), in the energy range above the intersection point of the two PESs. As a result, high-energy phonon states contribute more strongly to the effective phonon energy $\bar{E}_\text{eff}(T)$ in Eq.~(\ref{average_phonon}), leading to a relatively high $\bar{E}_\text{eff}(T)$. According to Eq. (\ref{MP_derivative}), this yields a large temperature derivative of $\ln \sigma_n^{\text{NMP}}$, so $\sigma_n^{\text{NMP}}$ varies by more than 12 orders of magnitude as $T$ increases from 0 to 600 K, as shown in Fig.~\ref{fig2}(d).

For the second defect with a smaller $\Delta Q$ [Fig. \ref{fig2}(b)], the two PESs are closer with their intersection on the same side (Marcus inverted region \cite{Marcus1965JCP}). Therefore, the wavefunctions of the initial-state and final-state phonons strongly overlap even for low-energy phonons (small $m$). In this case, low-energy phonon states contribute more to $\bar{E}_\text{eff}(T)$, leading to a lower $\bar{E}_\text{eff}(T)$ and a smaller $T$ derivative of $\ln \sigma_n^{\text{NMP}}$ than in the first defect with large lattice relaxation. The corresponding $\sigma_n^{\text{NMP}}$ changes slightly with $T$, as shown in Fig. \ref{fig2}(e).


As shown above, the transition is dominated by a selected subset of phonon states, which we refer to as the effective phonon states that determine $\bar{E}_\text{eff}(T)$. For defects with different $\omega$, $\Delta Q$, and $\Delta E$, both the thermal occupations $\rho_m$ and the phonon wavefunction overlaps $L_m$ differ, so the selected phonon states also differ.
This phonon-selection mechanism is entirely absent in the Henry-Lang model which uses a constant $E_\text{b}$, so $\sigma_n^\text{HL}$ fails to capture the difference in $\Delta Q$. The discrepancy between $\sigma_n^\text{HL}$ and $\sigma_n^\text{NMP}$ is relatively small for the first defect with large $\Delta Q$  [Fig. \ref{fig2}(d)], while it becomes more pronounced for the second defect with small $\Delta Q$ [Fig. \ref{fig2}(e)].

\subsection{Application of NMP model in DLTS analysis}\label{subsec2-3}

\begin{figure}[htbp]
	\centering
	\includegraphics[width=0.9\textwidth]{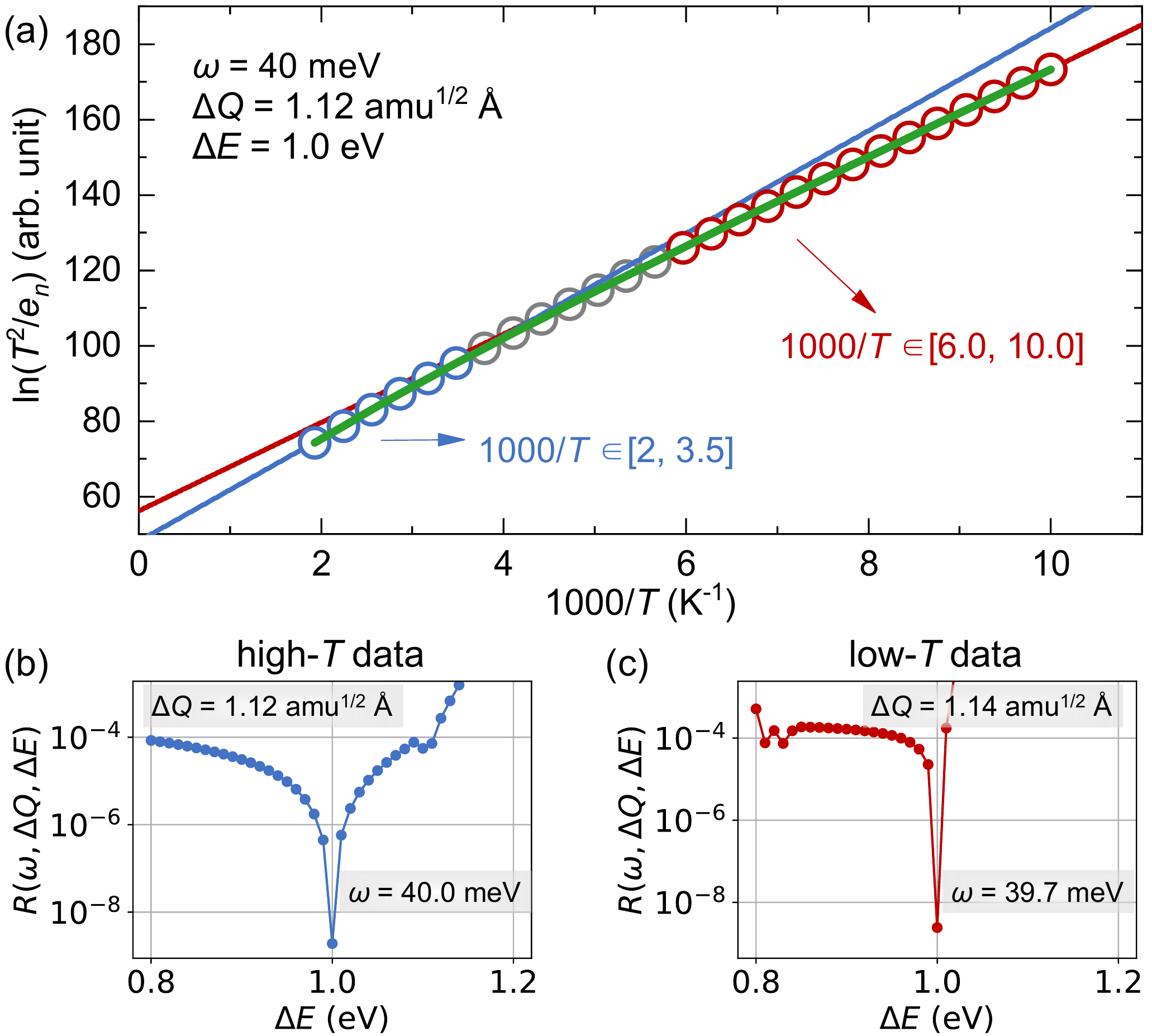}
	\caption{(a) Exact (open circles) and fitted (lines) $\ln(T^2/e_n)$ as functions of $1000/T$ for a model defect system. The blue and red lines show the linear fitting of the high-$T$ and low-$T$ data. The two green lines overlap, which show the fitting using Eq. (\ref{DLTS_refined}). (b, c) Minimized relative-residual function $R$ at different $\Delta E$ when (b) high-$T$ and (c) low-$T$ data are used for fitting. } 
	\label{fig3}
\end{figure}

Now we introduce how to use our NMP model for fitting the $e_n$ vs. $T$ ($\sigma_n$ vs. $T$) data measured by DLTS. Since Eqs. (\ref{multiphonon}-\ref{MP_integration}) indicate that $\sigma_n^{\text{NMP}}(T)$ depends also on $\omega,\Delta Q,\Delta E$, we define a relative-residual function in the DLTS temperature window [$T_1,T_2$],
\begin{equation}\label{DLTS_refined}
\begin{aligned}
    &R(\omega, \Delta Q, \Delta E)=\int_{T_1}^{T_2} \left[1-\frac{\sigma_n^{\text{NMP}}(T;\omega,\Delta Q,\Delta E)}{\sigma_n^{\text{exp}}(T;\Delta E)} \right]^2 dT,
\end{aligned}
\end{equation}
where $\sigma_n^{\text{NMP}}(T;\omega,\Delta Q,\Delta E)$ is the theoretical $\sigma_n$ calculated via Eq. (\ref{MP_integration}), and $\sigma_n^{\text{exp}}(T;\Delta E)$ is the experimental $\sigma_n$, which can be derived from $e_n$ measured by DLTS,  
    $\sigma_n^{\text{exp}}(T;\Delta E)=e_n \gamma^{-1} T^{-2}\exp[\Delta E/(k_BT)]$.
$\omega,\Delta Q,\Delta E$ are taken as the fitting parameters for minimizing the relative-residual function. 

To test whether fitting the relative-residual function defined here can give the correct $T$-dependence of $e_n$ and $\sigma_n$, we use the defect in Fig. \ref{fig2}(b) as a model system. Since its $\omega,\Delta Q,\Delta E$ are known, its $\sigma_n$, $e_n$ and $\ln(T^2/e_n)$ in $T=100-500$ K can be calculated exactly according to Eq. (\ref{MP_integration}). These exact values are then treated as experimental data, as shown by the circles in Fig. \ref{fig3}(a). We first fit the $\ln(T^2/e_n)$ vs. $1000/T$ relation in the $T$ range 285-500 K (blue circles) using our NMP model, and find the fitted green line passes precisely through all the exact data (circles),  not only within the fitting range but also outside it, down to 100 K. We then repeat the fitting using only the low-$T$ data in $T=100-167$ K, and again obtain excellent agreement over the entire $T=100$-$500$ K range. The $\omega$, $\Delta Q$, and $\Delta E$ fitted from the high- and low-$T$ data are nearly identical [Figs. \ref{fig3}(b) and \ref{fig3}(c)] and are also consistent with the original values in Fig. \ref{fig3}(a). These results demonstrate that the defined relative-residual function provides an efficient and reliable way to fit the $T$ dependence of $e_n$ and $\sigma_n$. A closer inspection of the circles in Fig. \ref{fig3}(a) shows that $\ln(T^2/e_n)$ is not strictly linear with $1000/T$ over the entire range $T=100$-$500$ K. This non-linearity is described correctly by our NMP model, but it cannot be described within the Henry-Lang model, which assumes a linear relation between $\ln(T^2/e_n)$ and $1000/T$. The blue line in Fig. \ref{fig3}(a) shows the linear fit of the high-$T$ data, while the red line shows that of the low-$T$ data. The two lines clearly do not overlap with each other and the fitted parameters ($\Delta E + E_\text{b}$ and $\sigma_{n0}$) also differ significantly. Due to different slopes of these two lines, the high-$T$ linear fit gives $\Delta E + E_\text{b}=1.17$ eV while the low-$T$ linear fit gives $\Delta E + E_\text{b}=1.01$ eV. Due to different intercepts at $1000/T=0$, the fitted carrier capture cross sections differ by more than 3 orders of magnitude. Therefore, the Henry-Lang model introduces serious arbitrary errors, which may cause controversies in the analysis of DLTS experimental results and pose a serious limit to the application of DLTS.

\subsection{Revisiting DLTS analysis of E2* trap in $\beta$-Ga$_2$O$_3$}\label{subsec2-4}

To test the application of our NMP model in real materials, we use it to analyze the DLTS data of the well-known E2* defect in $\beta$-Ga$_2$O$_3$, which was identified as an important deep trap by a series of experiments \cite{Mcglone,Ingebrigtsen2018APL, Zhang2024APL, Langorgen2024JAP,Dawe2025APL-M}. The $e_n$ vs. $T$ data measured by DLTS had been analyzed using the constant-$\sigma_n$ model and Henry-Lang model, giving $\sigma_n$ of $10^{-14}$-$10^{-15}$ cm$^{2}$, and $\Delta E$ of $0.70$-$0.75$ eV \cite{Mcglone,Ingebrigtsen2018APL, Zhang2024APL, Langorgen2024JAP,Dawe2025APL-M}. In the following, we will show huge arbitrary errors exist in such analysis, and then use the NMP model to analyze the same data \cite{Mcglone}.

\begin{figure}[htbp]
	\centering
	\includegraphics[width=0.9\textwidth]{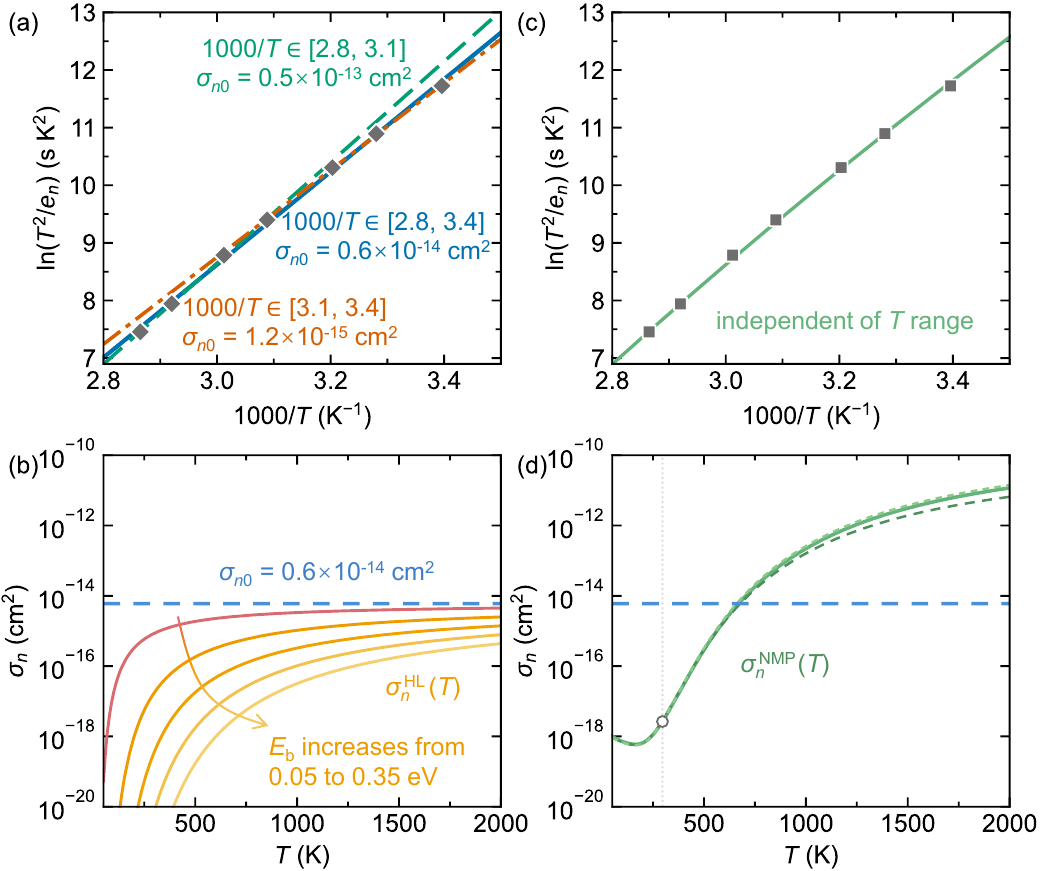}
	\caption{Comparison between the existing models and NMP model on describing E2* trap in $\beta$-Ga$_2$O$_3$. (a) Linear fit of $\ln(T^2/e_n)$ vs $1000/T$, and (b) calculated $\sigma_n$ as functions of $T$ ($\Delta E + E_b$ is fixed, while $E_\text{b}$ varies from 0.05 to 0.35 eV), using the constant-$\sigma_n$ model and Henry-Lang model. (c) Nonlinear fit of $\ln(T^2/e_n)$ vs $1000/T$, and (d) calculated $\sigma_n$ vs. $T$, according to the rigorous NMP model.}
	\label{fig4}
\end{figure}

Fig. \ref{fig4}(a) shows the measured $\ln(T^2/e_n)$ vs. $1000/T$ data (circles). As we can see, the data points deviate from an exact linear relation. However, in the analysis using the constant-$\sigma_n$ model and Henry-Lang model, they are fitted by a linear function, which introduces arbitrary errors. If we perform the linear fitting for the data in the $T=295-325$ K range, the fitted $\sigma_{n0}=1.2\times10^{-15}$ cm$^2$ and $\Delta E+E_\text{b}=0.65$ eV according to the Henry-Lang model in Eq. (\ref{Arrhenius}). However, the linear fitting in the $T=325-350$ K range gives very different results, $\sigma_{n0}=0.5\times10^{-13}$ cm$^2$ and $\Delta E+E_\text{b}=0.75$ eV; the fitting in the whole $T=295-350$ range gives $\sigma_{n0}=0.6\times10^{-14}$ cm$^2$ and $\Delta E+E_\text{b}=0.70$ eV. Therefore, the three lines show arbitrary errors in $\sigma_{n0}$ as large as 2 orders of magnitude. Furthermore, the fitting only yields the sum $\Delta E+E_\text{b}$, but the exact values of $\Delta E$ and $E_b$ are still unknown. Fig. \ref{fig4}(b) shows that when $E_\text{b}$ increases from 0 (corresponding to the constant-$\sigma_n$ model) to 0.35 eV, the $\sigma_n(T)$ curves change dramatically from the blue dashed line to the four orange color-gradient curves, and $\sigma_n$ changes from $10^{-14}$ to $10^{-20}$ cm$^2$ at room temperature, indicating an arbitrary error of 6 orders of magnitude caused by the uncertainty in $\Delta E$ and $E_\text{b}$. Such an uncertainty poses a serious limit to the application of Henry-Lang model, which makes the constant-$\sigma_n$ model even more widely used in experiments, despite its obviously unreasonable assumption of a temperature-independent $\sigma_n$.

Our NMP model solves the arbitrary error issue in the constant-$\sigma_n$ model and Henry-Lang model. When we perform the fitting using our defined relative-residual function in the $T=295-325$ K range, the fitted $\omega=112$ meV, $\Delta Q=1.66$ amu$^{1/2}$ $\text{\AA}$, $\Delta E=0.47$ eV; these values change slightly to $\omega=115$ meV, $\Delta Q=1.68$ amu$^{1/2}$ $\text{\AA}$, $\Delta E=0.52$ eV if the data in the $T=325-350$ K range are used for fitting, and change also slightly to $\omega=116$ meV, $\Delta Q=1.66$ amu$^{1/2}$ $\text{\AA}$, $\Delta E=0.50$ eV if the fitting is performed in the whole $T=295-350$ K range. As shown in Fig. \ref{fig4}(c), all the three fitting curves converge to a single green curve, which passes through the original circles, indicating that our NMP model does not show arbitrary errors. The fitted $\Delta E=0.5$ eV deviates by 0.2 eV from the value reported in Ref. \cite{Mcglone}, and the calculated $\sigma_n(T)$ curve in Fig. \ref{fig4}(d) also shows a very different behavior compared to those from the Henry-Lang model in Fig. \ref{fig4}(b). At room temperature, $\sigma_n$ is $2.9 \times 10^{-18}$ cm$^2$, nearly four orders of magnitude smaller than the value obtained by the Henry-Lang model (when $E_\text{b}=0$, corresponding to the constant-$\sigma_n$ model) \cite{Mcglone}. This means the impact of this E2* defect in Ga$_2$O$_3$-based devices could be significantly overestimated by the existing models.

\section{Discussion}\label{sec3}

\begin{table}[h]
\caption{DLTS fitted properties of 24 defects in 12 semiconductors using the Henry-Lang model (linear fit) and the NMP model (nonlinear fit). The linear fit of the Henry-Lang model gives $\Delta E+ E_\text{b}$ (in eV) and apparent carrier capture cross section $\sigma_{n0}$ (in cm$^2$) of Eq. (\ref{Arrhenius}). The nonlinear fit of the NMP model gives $\omega$, $\Delta Q$, $\Delta E$ of Eq. (\ref{DLTS_refined}), from which $E_\text{b}$ (in eV), $\sigma_n^{\text{NMP}}(T)$ (in cm$^2$) at different temperature $T$ can be derived.}\label{tab1}
\begin{tabular*}{\textwidth}{@{\extracolsep\fill}lccccccc}
\toprule%
& \multicolumn{2}{@{}c@{}}{Henry-Lang model (linear fit)} & \multicolumn{4}{@{}c@{}}{NMP model (nonlinear fit)} \\\cmidrule{2-3}\cmidrule{4-7}%
Defect & $\Delta E+ E_\text{b}$ & $\sigma_{n0}$ & $\Delta E$  & $E_\text{b}$ & $\sigma_n^{\text{NMP}}$ (300 K) & $\sigma_n^{\text{NMP}}$ (600 K)  \\
\midrule
Si: H2 \cite{Simoen_2019} & 0.31 & $0.8\times10^{-15}$ & 0.17 & 0.25 & $4.7\times10^{-18}$ & $1.9\times10^{-16}$ \\
Si: Pt \cite{Si-Pt}  & 0.22 & $1.3\times10^{-15}$  & 0.23 & 0.01 & $1.5\times10^{-15}$ &  $0.7\times10^{-15}$ \\
 &  &  &   & & \\
Ge: E$_{0.16}$ \cite{Coelho2013JAP}  & 0.16 & $1.2\times10^{-13}$  & 0.11 & 0.14 & $4.2\times10^{-13}$ &  $2.6\times10^{-12}$ \\
Ge: H$_{0.22}$ \cite{Coelho2013JAP}  & 0.22 & $2.2\times10^{-13}$  & 0.19 & 0.17 & $0.8\times10^{-12}$ &  $0.7\times10^{-11}$ \\
 &  &  &   & & \\
GaAs: EL2 \cite{thesis}  & 0.82 & $1.3\times10^{-13}$  & 0.78 & 0.09 & $3.0\times10^{-14}$ &  $0.6\times10^{-13}$ \\
GaAs: E$_{0.63}$ \cite{Tunhuma2016JAP}  & 0.69 & $2.1\times10^{-13}$  & 0.34 & 0.71 & $2.8\times10^{-19}$ &  $0.7\times10^{-15}$ \\
 &  &  &   & & \\
GaN: E \cite{Yang2020EDL} & 0.86 & $0.6\times10^{-14}$ & 0.77 & 0.94 & $1.5\times10^{-15}$ & $1.1\times10^{-13}$ \\
GaN: H \cite{Yang2020EDL} &  0.48 & $3.2\times10^{-19}$ & 0.21 & 0.82 & $0.8\times10^{-23}$ & $1.5\times10^{-20}$ \\
 &  &  &   & & \\
InP: H4 \cite{InP-2010} &  0.30 & $4.2\times10^{-17}$ & 0.17 & 0.62 & $0.6\times10^{-18}$ & $0.8\times10^{-15}$ \\
InP: H5 \cite{InP-2010} &  0.54 & $0.8\times10^{-14}$ & 0.28 & 0.60 & $2.5\times10^{-19}$ & $3.9\times10^{-16}$ \\
 &  &  &   & & \\
4H-SiC: S1 \cite{bathen2019npj} &  0.42 & $0.7\times10^{-14}$ & 0.31 & 0.58 & $0.7\times10^{-15}$ & $0.8\times10^{-12}$ \\
4H-SiC: S2 \cite{David2004JAP} &  0.67 & $1.4\times10^{-15}$ & 0.50 & 0.37 & $1.9\times10^{-18}$ & $0.8\times10^{-16}$ \\
 &  &  &   & & \\
Ga$_2$O$_3$: E1 \cite{Irmscher2011JAP} &  0.54 & $2.0\times10^{-14}$ & 0.57 & 0.01 & $0.7\times10^{-13}$ & $3.4\times10^{-14}$ \\
Ga$_2$O$_3$: E2* \cite{Mcglone} &  0.70 & $0.6\times10^{-14}$ & 0.50 & 0.87 & $2.9\times10^{-18}$ & $2.1\times10^{-15}$ \\
 &  &  &   & & \\
CdTe: H$_{0.26}$ \cite{CdTe1997} &  0.26 & $1.3\times10^{-17}$ & 0.15 & 0.59 & $0.6\times10^{-18}$ & $0.7\times10^{-15}$ \\
CdTe: E$_{0.49}$ \cite{CdTe1997} &  0.49 & $1.8\times10^{-13}$ & 0.33 & 0.67 & $4.1\times10^{-16}$ & $0.6\times10^{-12}$ \\
 &  &  &   & & \\
ZnO: E3 \cite{ZnO-E3} &  0.31 & $2.0\times10^{-15}$ & 0.29 & 0.27 & $1.8\times10^{-15}$ & $2.4\times10^{-14}$ \\
ZnO: c \cite{ZnO-c} &  0.18 & $0.6\times10^{-16}$ & 0.12 & 0.47 & $0.9\times10^{-15}$ & $0.6\times10^{-12}$ \\
 &  &  &   & & \\
Sb$_2$Se$_3$: E1 \cite{wen2018NC}  & 0.61 & $1.7\times10^{-12}$ & 0.31 & 0.59 & $1.5\times10^{-17}$ & $4.2\times10^{-14}$ \\
Sb$_2$Se$_3$: H1 \cite{wen2018NC} & 0.49 & $0.6\times10^{-17}$ & 0.48 & 0.07 &  $4.1\times10^{-18}$ & $0.8\times10^{-17}$ \\
 &  &  &   & & \\
CIGS: A2 \cite{Urbaniak2016JPCM} & 0.25 & $1.3\times10^{-17}$ & 0.12 & 0.49 &  $4.6\times10^{-19}$ & $3.6\times10^{-16}$ \\
CIGS: A3 \cite{Urbaniak2016JPCM} & 0.09 & $3.9\times10^{-19}$ & 0.10 & 0.01 &  $0.6\times10^{-18}$ & $3.1\times10^{-19}$ \\
 &  &  &   & & \\
CZTS: E$_\text{A2}$ \cite{Das2014APL} & 0.32 & $0.6\times10^{-17}$ & 0.26 & 0.39 &  $0.5\times10^{-18}$ & $1.2\times10^{-17}$  \\
CZTS: H3\cite{CZTS-H3} & 0.43 & $3.0\times10^{-16}$ & 0.27 & 0.71 &  $0.7\times10^{-18}$ & $0.9\times10^{-15}$  \\

\botrule
\end{tabular*}
\end{table}

The case of the E2* trap in $\beta$-Ga$_2$O$_3$ demonstrates the large errors in the DLTS analysis using existing models, compared to the unambiguous analysis using the rigorous NMP model. Such large errors may exist generally in many previous DLTS studies on defects in a large number of different semiconductors.
We collected the DLTS measured $\ln(T^2/e_n)$ vs. $1000/T$ data for 24 different defects in 12 semiconductors as reported in literature \cite{Simoen_2019,Si-Pt, Coelho2013JAP,thesis, Tunhuma2016JAP,Yang2020EDL,InP-2010,bathen2019npj,David2004JAP,Irmscher2011JAP,Mcglone,CdTe1997,ZnO-E3,ZnO-c,wen2018NC,Urbaniak2016JPCM,Das2014APL,CZTS-H3}, and compared the defect properties fitted using the Henry-Lang model (linear fit) and NMP model, as listed in Table \ref{tab1}. 

In the Henry-Lang model, the linear fitting gives $\Delta E+ E_\text{b}$ and $\sigma_{n0}$, and the temperature dependent $\sigma_n^{\text{HL}}(T)$ can be derived according to $\sigma_n^{\text{HL}}(T)= T^{-1}\sigma_{n0} \exp \left(-E_\text{b}/k_\text{B}T \right)$ in principle. However, since only the sum $\Delta E+ E_\text{b}$ is obtained from the fitting while the specific value of $E_\text{b}$ is still unknown, $\sigma_n^{\text{HL}}(T)$ at the working temperature $T$ of a semiconductor device cannot be determined and used for evaluating the influence of defects on carrier recombination (lifetime), charge trapping, and device reliability. Instead, the apparent carrier capture cross section $\sigma_{n0}$ is often used directly for these purposes. Although it is obviously unreasonable, but it is very common in the carrier dynamics studies based on the DLTS defect characterization nowadays \cite{Polyakov2018APL,lian2021NC, Chen2024PRM}. For most of the defects, Table \ref{tab1} shows that $\sigma_{n0}$ fitted using the Henry-Lang model differs by multiple orders of magnitude from $\sigma_n^{\text{NMP}}(T)$ at $T=300$ K fitted using the rigorous NMP model for the same DLTS measured $\ln(T^2/e_n)$ vs. $1000/T$ data. For instance, for E$_{0.63}$ defect in GaAs \cite{Tunhuma2016JAP}, H5 defect in InP \cite{InP-2010}, and E1 defect in Sb$_2$Se$_3$ \cite{wen2018NC}, $\sigma_{n0}$ is overestimated by approximately 5 to 6 orders of magnitude compared to $\sigma_n^{\text{NMP}}$ (300 K). Differences over 3 orders of magnitude are found for many other defects, such as H2 in Si \cite{Simoen_2019}, H in GaN \cite{Yang2020EDL}, S2 in SiC \cite{David2004JAP}, and E$_{0.49}$ in CdTe \cite{CdTe1997}. The differences between $\sigma_{n0}$ and $\sigma_n^{\text{NMP}}$(300 K) are small for only 3 cases, including Pt dopant in Si \cite{Si-Pt}, E1 defect in Ga$_2$O$_3$ \cite{Irmscher2011JAP}, and A3 defect in Cu(In,Ga)Se$_2$ \cite{Urbaniak2016JPCM}. In these cases, the classical barrier $E_\text{b}$ is nearly negligible, leading to a weak temperature dependence of $\sigma_n(T)$, which explains why the Henry-Lang model and NMP model gives small differences between $\sigma_{n0}$, $\sigma_n^{\text{NMP}}$ (300 K), and $\sigma_n^{\text{NMP}}$ (600 K). 

Unfortunately, $E_\text{b}$ is large for most of the defects in Table \ref{tab1}, so $\sigma_{n0}$ is usually seriously overestimated relative to the correct $\sigma_n^{\text{NMP}}(T)$ at $T=300$ K. When this $\sigma_{n0}$ is used as the carrier capture cross sections in the carrier dynamic studies and device simulation or design, it can cause serious errors in the derived carrier capture rate and device performance at the working temperature (e.g., $T=300$ K), causing incorrect assessment on the roles of these defects in device operation. Such errors in the semiconductor defect studies have not yet been assessed until the present study with the rigorous NMP model. Therefore, we call for comprehensive revisiting studies on the carrier capture cross sections and energy levels of defects that are reported based on the Henry-Lang model fitting of DLTS measured $\ln(T^2/e_n)$ vs. $1000/T$ data.

\begin{figure}[htbp]
	\centering
	\includegraphics[width=0.9\textwidth]{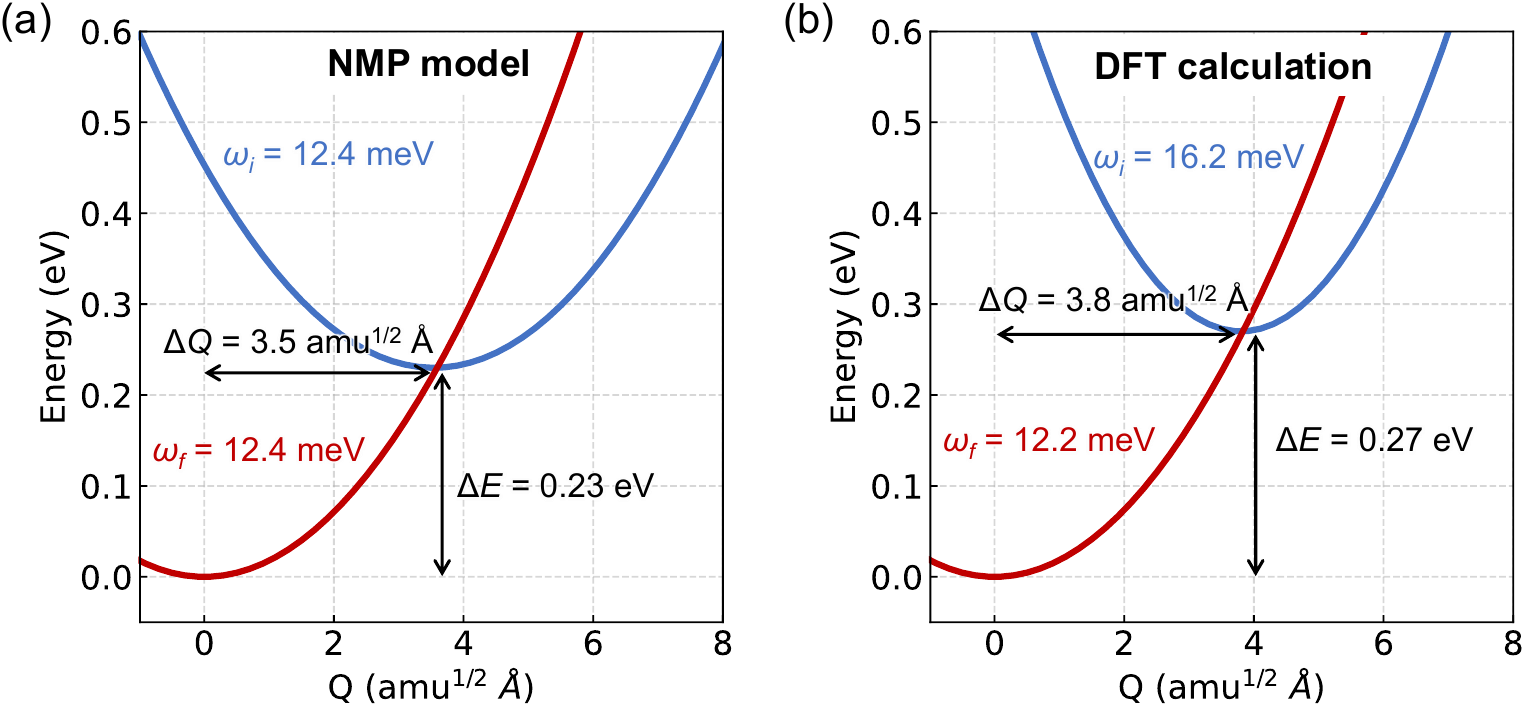}
	\caption{Configuration coordinate diagrams of electron carrier capture at Pt$_\text{Si}$ in Si, constructed from (a) the NMP model fitted ($\omega$, $\Delta Q$, $\Delta E$) parameters of the DLTS experimental data \cite{Si-Pt}, and (b) first-principles calculation of $(0/-1)$ transition of Pt$_\text{Si}$ \cite{daike}.} 
	\label{fig5}
\end{figure}

Since our present study shows that the NMP model should be adopted in the DLTS analysis, we have developed a DLTS NMP fitting platform on the website https://www.semidefect.com/dlts. The platform (free access after registration) enables users to reproduce the results reported in Table \ref{tab1}, and also to apply the NMP model for analyzing their own DLTS data. We expect this platform can become a standard analysis tool for the future DLTS studies. 

To test whether the NMP model fitting of DLTS experimental data can give the defect properties that are consistent with first-principles calculations, we take the Si: Pt defect (Pt dopant in crystalline silicon) as an example. This defect is a good example system, because the Pt dopant is well-known to substitute the Si atom (taking the Pt$_\text{Si}$ structural configuration) and density functional theory (DFT) calculations have shown clearly that Pt$_\text{Si}$ can capture an electron from the conduction band through the (0/-1) transition with $\Delta E=0.27$ eV, $\omega_i=16.2$ meV, $\omega_f=12.2$ meV, $\Delta Q=3.8$ amu$^{1/2}$ \AA  \cite{daike}, as shown by the DFT calculated configuration coordinate diagram in Fig. \ref{fig5}(b). Interestingly, our NMP model fitting of the DLTS measured $\ln(T^2/e_n)$ vs. $1000/T$ data shows $\omega=12.4$ meV, $\Delta Q=3.5$ amu$^{1/2}$ \AA, $\Delta E=0.23$ eV (Fig. \ref{fig5}(a)), in good agreement with the DFT calculated values for the $(0/-1)$ transition of Pt$_\text{Si}$, indicating that the DLTS signal should originate from this transition level. The good agreement also demonstrates that our NMP model describe the nonradiative multiphonon transition correctly.

\section{Conclusion}\label{sec13}

\par In summary, we show the widely-used Henry-Lang model and constant-$\sigma_n$ model fail in describing the temperature dependence of carrier capture cross section, and can cause large arbitrary errors when fitting the DLTS observed emission rate. We develop a rigorous NMP model, which can be used to perform the unambiguous fitting of DLTS experiments and derive the correct energy level and carrier capture cross section of defects at different temperatures. We apply the NMP model to fit the DLTS experimental data reported in literature for 21 different defects in 12 semiconductors, and demonstrate  that the fitting using the Henry-Lang model overestimates the carrier capture cross sections by multiple orders of magnitude at room temperature, causing serious errors in the previous defect studies using DLTS. We propose that the NMP model should be adopted and replace the existing models in the future DLTS defect characterization and defect physics study.

\section{Data availability}
The source data for reanlyzing the DLTS results can be found in the cited literature.

\section{Code availability}
The implementation of the NMP fit is made available at https://www.semidefect.com/dlts.

\bmhead{Supplementary information}

Supplementary Information.

\bmhead{Acknowledgements}

This work was supported by National Key Research and Development Program of China (2024YFB4205002), National Natural Science Foundation of China (12334005, 12188101 and 12404089), Science and Technology Commission of Shanghai Municipality (24JD1400600).


\bibliography{sn-bibliography}

\end{document}